\documentclass[pre,twocolumn,twoside,showpacs,superscriptaddress]{revtex4}
\usepackage{graphicx,graphics,amssymb,amsmath,epsf,color}
\epsfclipon

\usepackage{color}
\usepackage{xcolor}

\newcommand{\rev}{\textcolor{black}}

\begin{document}
\title{Stable and flexible system for glucose homeostasis}
\author{Hyunsuk Hong}
\affiliation{Department of Physics and Research Institute of Physics and Chemistry, Chonbuk National University, Jeonju 561-756, Korea}
\author{Junghyo Jo}
\email{jojunghyo@apctp.org}
\affiliation{Asia Pacific Center for Theoretical Physics, Pohang, Korea}
\affiliation{Department of Physics, POSTECH, Pohang, Korea}
\author{Sang-Jin Sin}
\affiliation{Department of Physics, Hanyang University, Seoul, Korea}

\date{\today}
\begin{abstract}
Pancreatic islets, controlling glucose homeostasis, consist of $\alpha$, $\beta$, 
and $\delta$ cells. It has been observed that $\alpha$ and $\beta$ cells 
generate out-of-phase synchronization in the release of glucagon and insulin, 
counter-regulatory hormones for increasing and decreasing glucose levels,
while $\beta$ and $\delta$ cells produce in-phase synchronization in the 
release of the insulin and somatostatin.
Pieces of interactions between the islet cells have been observed for 
a long time, although their physiological role as a whole has not been explored yet.
We model the synchronized hormone pulses of islets with coupled phase oscillators
that incorporate the observed cellular interactions.
The integrated model shows that the interaction from $\beta$ to $\delta$ cells,
of which sign has controversial reports, should be positive 
to reproduce the in-phase synchronization between $\beta$ and $\delta$ cells.
The model also suggests that $\delta$ cells  
help the islet system flexibly respond to changes of glucose environment.
\end{abstract}

\maketitle

\section{Introduction}
Life maintain energy through metabolism.
Among the two major fuels of our body, glucose and lipid, 
glucose is the primary energy source, particularly for brain cells.
Therefore, maintaining glucose levels constant, {\it glucose homeostasis}, is
essential for life. Its failure leads to a metabolic disease, diabetes.
Islets of Langerhans in the pancreas play
a critical role for maintaing the glucose homeostasis.
It is composed of three major cell types: $\alpha$, $\beta$, and $\delta$ cells.
During fasting and fed states, $\alpha$ and $\beta$ cells secrete glucagon and insulin, respectively,
for increasing and decreasing glucose levels.
At first sight, these two reciprocal cells seem
sufficient for controlling glucose levels.
However, a third one, $\delta$ cell, has been found,
and its role on the glucose homeostasis has yet to be unveiled.

Like other hormones in the body,
the insulin and glucagon secretions show rhythmic behavior~\cite{ref:Lefebvre}. 
Their oscillation with $5-10$ minute periods
have  been repeatedly observed not only in the cells within 
islets~\cite{ref:Bergsten}, but also in isolated cells~\cite{ref:Grapengiesser}.
\rev{In particular, the periodic insulin release has been extensively studied 
with mathematical modeling~\cite{ref:Bertram}.}
It has been reported that glucagon and insulin 
exhibit out-of-phase synchronization
both in vivo~\cite{ref:Menge} and in vitro~\cite{ref:Hellman}. 
The in vitro study~\cite{ref:Hellman} has also shown that insulin and somatostatin (secreted by $\delta$ cells)
have in-phase synchronization.
In addition, Menge et al demonstrated that the out-of-phase 
synchronization is disrupted in diabetes patients, suggesting the 
physiological importance of the coordinated
insulin and glucagon secretion~\cite{ref:Menge}.
It has long been observed that the endocrine cells 
interact with each other through hormones and/or 
neurotransmitters~\cite{ref:Koh}. 

\rev{
Synchronization between coupled oscillators has long been studied in physics~\cite{ref:synch}.
In particular, the Kuramoto model has been introduced to explain collective behavior
such as synchronization in the population of coupled oscillators~\cite{ref:Kuramoto},
and recently generalized by allowing the coupling with arbitrary phase shift ~\cite{ref:Pikovsky_Rosenblum}.
In other words, the general model can have arbitrary signs and strengths of coupling,
while the original model has only positive coupling.
Hong and Strogatz have proposed an interesting specification of the generalized Kuramoto model
in which two populations of conformists (having positive coupling) and contrarians (negative coupling)
interact and show rich dynamics~\cite{ref:Hong_Strogatz_PRL}.
As a natural extension, the synchronization between three symmetrically-distinct populations
is of particular interest. Here we introduce a perfect realization of the three-body interaction
in biology.
}

\rev{Using the generalized Kuramoto model,  we specifically} answer the following question:
Are the observed pieces of 
local interactions between $\alpha$, $\beta$, and $\delta$ 
cells sufficient and consistent to explain the 
synchronized hormone secretion?
We also explore the role of the third population, $\delta$ cells,
additional to the counter-regulating $\alpha$ and $\beta$ cells
in the control system for homeostasis.

This paper consists of five sections.
In Sec. II, synchronized hormone pulses of $\alpha$, $\beta$, and $\delta$ cells
are described by the three coupled phase oscillators. 
Section III derives a generalized islet model that considers population of each cell type.
Section IV presents results and predictions of the islet model.
Finally, Sec. V summarizes and discusses the results.

\section{Islet model}
To understand the synchronized hormone pulses 
in the pancreatic islets, we simply regard the endocrine cells as intrinsic 
oscillators producing pulsatile hormones 
because isolated cells still show oscillations in the absence of 
neighboring cells.
Then, the attractive or repulsive interaction between the oscillators
play a role to synchronize them in phase or out of phase.
Since we are interested in only the phases of the three interacting oscillators of
$\alpha$, $\beta$, and $\delta$ cells, their synchronization dynamics
can be described by three coupled oscillators
with Kuramoto-type interactions~\cite{ref:Kuramoto}:
\begin{eqnarray}
\label{eq:model1}
\dot{\theta}_1 &=& \omega_1 + J_{21} \sin(\theta_2-\theta_1) + J_{31} \sin(\theta_3-\theta_1),  \\
\dot{\theta}_2 &=& \omega_2 + J_{12} \sin(\theta_1-\theta_2) + J_{32} \sin(\theta_3-\theta_2),  \\
\dot{\theta}_3 &=& \omega_3 + J_{13} \sin(\theta_1-\theta_3) + J_{23} \sin(\theta_2-\theta_3).
\end{eqnarray}
The subscripts  1, 2, and 3 here correspond to $\alpha$, $\beta$, 
and $\delta$ cells, respectively. 

\begin{figure}
\includegraphics[width=7cm]{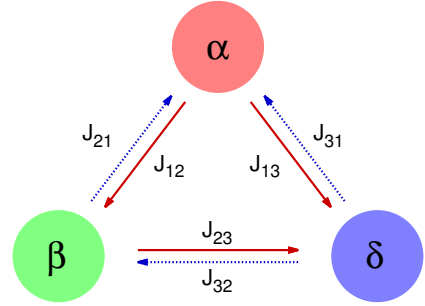}
\caption{(Color online)
Schematic diagram of $\alpha$, $\beta$, and $\delta$ cells in the pancreatic 
islets. 
The red (solid) arrows represent the attractive 
interaction from the $s$ cell 
to the $s^{\prime}$ cell, with the positive 
strength $(J_{s s^{\prime}}>0)$, 
and the blue (dotted) arrows denote the repulsive 
interaction, with 
the negative strength $(J_{s s^{\prime}}<0)$, respectively.  
The sign of cellular interactions has been taken from the known facts 
based on the observations (see the Table~\ref{table1}).} 
\label{fig:cell_diagram}
\end{figure}

The variable $\omega_{1,2,3}$ denotes their natural frequencies. 
$J_{s^{\prime} s}$ represents the coupling (interaction) strength from  
the $s^{\prime}$ cell onto the $s$ cell (Fig.~\ref{fig:cell_diagram}).
The sign of the couplings $J_{s^{\prime} s}$ between $\alpha$, $\beta$, and $\delta$ cells
can be found in the literatures and is summarized in the Table~\ref{table1}.
We consider here the case of asymmetric couplings
($J_{s^{\prime} s} \neq J_{s s^{\prime}}$), and further
include repulsive interaction with negative strength ($J_{s^{\prime} s}<0 $) 
in addition to the attractive one with positive value ($J_{s^{\prime} s} > 0$).
The repulsive and attractive coupling has also been known to appear 
in the neural networks with {\it excitatory} and 
{\it inhibitory} coupling~\cite{ref:excitatory_inhibitory}, where the positive 
coupling is for the excitatory neurons, and the negative one for the 
inhibitory neurons, respectively.  
To facilitate the comparison with the recent reports~\cite{ref:Hellman},
we suppose that $J_{12}=J_{13}=p~(>0) $, $-J_{21}=J_{23}=q~(>0)$, and 
$-J_{31}=-J_{32}=r~(>0)$ for the interactions between cell types. 
It is reasonable to assume that the interaction strengths 
from the $s$ cell to the  
$s^{\prime}$ and $s^{\prime \prime}$ cell are equivalent as
$|J_{ss^{\prime}}|=|J_{ss^{\prime \prime}}|$,
because the interactions are realized by the same molecules
secreted from the $s$ cell.

To simplify our system, and in reasonable agreement with observations~\cite{ref:Grapengiesser},
we assume that the oscillators in Eq.~(\ref{eq:model}) have the same natural
frequency ($\omega_{1,2,3}=\omega$).
Then, since we are interested in the phase differences between cell types,
Eq.~(\ref{eq:model1}) is then reduced to 
\begin{eqnarray}
\label{eq:u1}
\dot{u} &=& (q - p)\sin u + r \big[\sin v + \sin (u-v) \big],  \\
\label{eq:v1}
\dot{v} &=& q \big[\sin u + \sin (u-v) \big] + (r - p)\sin v,
\end{eqnarray}
where $u \equiv \theta_1 - \theta_2$ and $v \equiv \theta_1 - \theta_3$.

In the following section, we derive a generalized islet model considering 
populations of each cell type in the islet. 
However, because the population model results in essentially the same conclusion,
readers who are not interested in the sophisticated analysis may skip Sec III.

\begin{table}
\begin{center}
\begin{tabular}{|c|c|c|c|} \hline
Parameter & Interaction & Sign & Reference \\ \hline\hline
$J_{12}$ & $\alpha \rightarrow \beta$ & $+$ & \cite{ref:J12} \\ \hline
$J_{13}$ & $\alpha \rightarrow \delta$ & $+$ & \cite{ref:J13} \\ \hline
$J_{21}$ & $\beta \rightarrow \alpha$ & $-$ & \cite{ref:J21} \\ \hline
$J_{23}$ & $\beta \rightarrow \delta$ & $+$ &  \cite{ref:J23a}\\ 
 &  & $0$ & \cite{ref:J23b} \\ \hline
$J_{31}$ & $\delta \rightarrow \alpha$ & $-$ & \cite{ref:J31J32} \\ \hline
$J_{32}$ & $\delta \rightarrow \beta$ & $-$ & \cite{ref:J31J32} \\ \hline
\end{tabular}
\caption[Table 1]{
Signs of cellular interactions in literatures.
} 
\label{table1}
\end{center}
\end{table}

\section{Population model}
Considering populations of each cell type, we develop a model of coupled phase oscillators for the cells 
in the islet which is governed by 
\begin{equation}
\label{eq:model}
\dot\phi^s_j = \omega^s_j + \frac{1}{N}\sum_{s^{\prime}=1}^{3}\sum_{k=1}^{N_{s^{\prime}}} J_{{s^{\prime}} s} 
\sin(\phi^{s^{\prime}}_k - \phi^s_j)
\end{equation}
{\color{black}{for $s=1,2,3$}}, where $j=1,\cdots, N_s$, and 
$\phi^s_j$ represents the phase/angle of the oscillator $j$ in 
subpopulation $s$.
The number $N_s$ is the size of the subpopulation $s$: 
$N=\sum_{s=1}^3 N_s$.   
The subpopulation with $s=1, 2, 3$ here corresponds to the subgroup 
that consists of $\alpha$, $\beta$, 
and $\delta$ cells, respectively. 
The variable $\omega^s_j$ denotes the natural frequency of the oscillator $j$ in 
the subpopulation $s$,
where we assumed that the oscillators have the same natural
frequency ($\omega^s_j=\omega$).
$J_{s^{\prime} s}$ represents the coupling (interaction) strength from the 
oscillators in the subpopulation $s^{\prime}$ onto those in the subpopulation 
$s$.  
\rev{We note that a model similar to Eq.~(\ref{eq:model}) has been 
introduced in previous studies~\cite{ref:Daniel_Steve,ref:Pikovsky_Rosenblum}.
}

Here we can set $\omega$ to zero without
loss of generality by the phase transformation:
$\phi^s_j \rightarrow \phi^s_j + \omega t$.
Eq.~(\ref{eq:model}) is then rewritten as 
\begin{equation}
\label{eq:phi_gh}
\dot\phi^s_j = g_s e^{i\phi^s_j} + {\bar g_s}e^{-i\phi^s_j},
\end{equation}
where 
$g_s=\frac{i}{2N}\sum_{s^{\prime}=1}^{3}\sum_{k=1}^{N_{s^{\prime}}}
J_{s^{\prime} s} e^{-i \phi^{s^{\prime}}_k}$
and $\bar g_s$ is its complex conjugate. 

Collective synchronization in the system of coupled oscillators is
conveniently measured by the complex order parameter~\cite{ref:Kuramoto,ref:synch}
\begin{equation}
\label{eq:Z}
Z\equiv R e^{i\Theta}=\frac{1}{N}\sum_{k=1}^{N}e^{i\phi_k},
\end{equation}
where $R$ is a {\it global} order parameter that measures 
the phase coherence over all oscillators for the whole system, 
and $\Theta$ is the average phase. 
This order parameter $Z$ can be {\color{black}{divided into three terms}}:  
\begin{equation}
\label{eq:z1_z2_z3}
Z(t)= n_1 z_1 + n_2 z_2 + n_3 z_3 
\end{equation}
with $n_s=N_s/N$ and $z_s=(1/N_s)\sum_{k=1}^{N_s} e^{i\phi^s_k}$, where 
$z_s$ represents a {\it local} order parameter for the 
subpopulation $s~(=1,2,3)$. 

We now consider {\color{black}{the continuum limit, $N\rightarrow \infty$}}.  
In this limit, the order 
parameter $Z(t)$ can be written as 
\begin{equation}
\label{eq:Z_f}
Z(t)=\int_0^{2\pi} e^{i\phi} f(\phi,t) d\phi,
\end{equation}
where $f(\phi,t)$ denotes the probability density {\color{black}{function}} of 
the phases that lie between $\phi$ and $\phi+d\phi$ at time $t$.
Following the Ott-Antonsen ansatz~\cite{ref:OA}, we let 
\begin{equation}
\label{eq:f}
f(\phi, t) = \frac{1}{2\pi} \Bigg\{1+\sum_{n=1}^{\infty} \big[\bar{\alpha}(t)^n e^{i n\phi} + \alpha(t)^n e^{-i n\phi} \big]\Bigg\}
\end{equation}
for some unknown function $\alpha$ that is independent of $\phi$.
We note that Eq.~(\ref{eq:f}) is 
equivalent to the usual form of the Poisson 
kernel~\cite{ref:Mobius_transformation}
\begin{equation}
\label{eq:f_Poisson}
f(\phi)=\frac{1}{2\pi}\frac{1-\rho^2}{1-2\rho\cos(\phi-\theta)+\rho^2},
\end{equation}
where $\rho$ and $\theta$ are defined via $\alpha=\rho e^{i\theta}$,
and $\sum_{n=1}^{\infty} \bar{\alpha}^n e^{i n \phi} = \bar{\alpha} e^{i \phi}/(1-\bar{\alpha} e^{i \phi})$ is used.
With this, we find that $\alpha(t)$ in Eq.~(\ref{eq:f}) can be interpreted 
as the order parameter $Z(t)$, 
where $\rho$ and $\theta$ correspond to $R$ and $\Theta$ in 
Eq.~(\ref{eq:Z}), respectively. 
On the Poisson submanifold that is expressed by Eq.~(\ref{eq:f}) 
each probability density function $f_s$ for the subpopulation $s$ is also 
a Poisson kernel, therefore it has the same Fourier 
expansion as Eq.~(\ref{eq:f}), with $\alpha_s$ instead of $\alpha$: 
$f_s = \frac{1}{2\pi} \{1+\sum_{n=1}^{\infty}[\bar{\alpha}_s(t)^n e^{i n\phi} + \alpha_s(t)^n e^{-i n\phi} ] \}$.
Here, $\alpha_s$ corresponds to the {\it local} order parameter 
for the subpopulation $s$: $\alpha_s=\rho_s e^{i\theta_s}=z_s$.  
Substituting Eq.~(\ref{eq:f}) into Eq.~(\ref{eq:Z_f}), 
we find $Z(t)=\alpha(t)$, which {\color{black}{further}} yields 
$Z(t)=\sum_{s=1}^{3}n_s\alpha_s(t)$.

Meanwhile, we expect that
the continuity equation is satisfied for each subpopulation as
\begin{equation}
\label{eq:continuity}
\frac{\partial f_s}{\partial t} + \frac{\partial}{\partial \phi} (f_s v_s) =0,
\end{equation} 
where $f_s$ is the probability density function for 
the subpopulation $s$, and $v_s$ is the velocity field given 
by $v_s (\phi, t)= g_s e^{i\phi} + \bar{g_s} e^{-i\phi}$.
Substituting $f_s$ and $v_s$ into Eq.~(\ref{eq:continuity}), we obtain 
\begin{equation}
\label{eq:alpha}
\big[\dot\alpha_s - i(g_s \alpha^2_s + \bar g_s) \big]
\sum_{n=1}^{\infty}n{\alpha_s}^{n-1} e^{-i n\phi} + c.c = 0,
\end{equation}
where $c.c$ denotes the complex conjugate of the first term. 
We find that the summation in Eq.~(\ref{eq:alpha}) does not vanish, 
thus the factor in front of the summation should be zero, which leads to 
\begin{equation}
\label{eq:alpha_s}
\dot\alpha_s = i(g_s \alpha^2_s + \bar g_s). 
\end{equation}
This means that $z_s(t)$ also evolves according to 
$\dot z_s = i(g_s z^2_s + \bar g_s)$. 

We supposed that $J_{12}=J_{13}=p~(>0) $, $-J_{21}=J_{23}=q~(>0)$, and 
$-J_{31}=-J_{32}=r~(>0)$ for the interactions between the subpopulations. 
For the subpopulation self-coupling, on the other hand,
we let $J_{11}=I_1$, $J_{22}=I_2$, and $J_{33}=I_3$, 
where $I_1, I_2, I_3$ are all larger than $p, q, r$, which means that 
the coupling strength within a group is stronger than that between the 
subpopulations.
With these interactions, and with the substitution of 
$\alpha_s=\rho_s e^{i\theta_s}$ into Eq.~(\ref{eq:alpha_s}) for $s=1,2,3$, 
we find that the dynamics of each subpopulation is governed by 
\begin{eqnarray}
\label{eq:rho1_theta1}
\dot{\rho_1} &=& \frac{1\!-\!\rho_1^2}{2} \big [I_1 n_1 \rho_1 - q n_2 \rho_2 
\cos u - r n_3 \rho_3 \cos v \big], \nonumber\\ 
\dot{\theta_1} &=& \frac{1\!+\!\rho_1^2}{2\rho_1} 
\big[ q n_2 \rho_2 \sin u + r n_3 \rho_3 \sin v \big],   \\
\label{eq:rho2_theta2}
\dot{\rho_2} &=& \frac{1\!-\!\rho_2^2}{2} \big[p n_1 \rho_1 \cos u 
+ I_2 n_2 \rho_2 -r n_3 \rho_3 \cos (u-v) \big], \nonumber \\
\dot{\theta_2} &=& \frac{1\!+\!\rho_2^2}{2\rho_2}  \big[p n_1 \rho_1 \sin u 
- r n_3 \rho_3 \sin (u-v) \big], \\
\label{eq:rho3_theta3}
\dot{\rho_3} &=& \frac{1\!-\!\rho_3^2}{2} \big[p n_1 \rho_1 \cos v 
+ q n_2 \rho_2 \cos(u-v)+I_3 n_3 \rho_3 \big], \nonumber \\
\dot{\theta_3} &=& \frac{1\!+\!\rho_3^2}{2\rho_3} \big[p n_1 \rho_1 \sin v 
- q n_2 \rho_2 \sin (u-v) \big],  
\end{eqnarray}
respectively, where $u\equiv \theta_1 - \theta_2$, and $v\equiv \theta_1 - \theta_3$. 
For one simple case, we can assume that each subpopulation is 
in {\it perfect} synchronization ($\rho_1=\rho_2=\rho_3=1$).   
We note that this assumption is 
consistent with the experimental observation
that $\beta$ cells, sharing gap-junction channels with adjacent
$\beta$ cells, are strongly synchronized ($\rho_2=1$)~\cite{ref:Ravier}.
\rev{The long-range interaction between remote $\beta$ cells has been
mechanically justified by showing that the gap junctions mediate calcium waves in islets
~\cite{ref:Benninger}.}
On the other hand, no clear evidence for self-synchronization of 
$\alpha$ and $\delta$ cells has been found.
However, pulsatile glucagon and somatostatin secretions
of $\alpha$ and $\delta$ cells imply their self-synchronization
($\rho_1=1$ and  $\rho_3=1$).
Otherwise, asynchronous hormone pulses would compensate each other,
and their averaged pulses would become flat.
Then, since we are interested in the phase differences between cell types,
Eq.~(\ref{eq:rho1_theta1})-(\ref{eq:rho3_theta3}) is then reduced to 
\begin{eqnarray}
\label{eq:u}
\dot{u} &=& (q n_2 - p n_1)\sin u + r n_3 \big[\sin v + \sin (u-v) \big],  \\
\label{eq:v}
\dot{v} &=& q n_2 \big[\sin u + \sin (u-v) \big] + (r n_3 - p n_1)\sin v.
\end{eqnarray}
Therefore, we arrived at the same conclusion in Eq.~(\ref{eq:u1})-(\ref{eq:v1}),
except for weighting population densities to the coupling strengths
($p \rightarrow p n_1$, $q \rightarrow q n_2$, and $r \rightarrow r n_3$).

\section{Model analysis}
We now examine the synchronization patterns of the islet model in Eq.~(\ref{eq:u1})-(\ref{eq:v1}).
Specifically, we pay attention to the fixed point $(u^*, v^*)$ that is 
obtained from $\dot u=0$ and $\dot v=0$. 
It is found that $(0, 0)$, $(0, \pm\pi)$, $(\pm\pi, 0)$, 
and $(\pm\pi, \pm\pi)$ are all 
fixed points of Eq.~(\ref{eq:u1}) and (\ref{eq:v1}).
Note that some parameter set ($p$, $q$, $r$) allows nontrivial fixed points ($u_{\pm}$, $v_{\pm}$),
satisfying $\tan u_{\pm} = \mp pqh/r(h^2-2pq)$ 
and $\tan v_{\pm} = \pm (p-q+r)h/(h^2-2pr)$
with $h=\sqrt{2pq+2qr+2rp-p^2-q^2-r^2}$.
The stability of the fixed points has been checked, using the 
linear stability analysis~\cite{footnote1}.

\begin{figure}
\includegraphics[width=9cm]{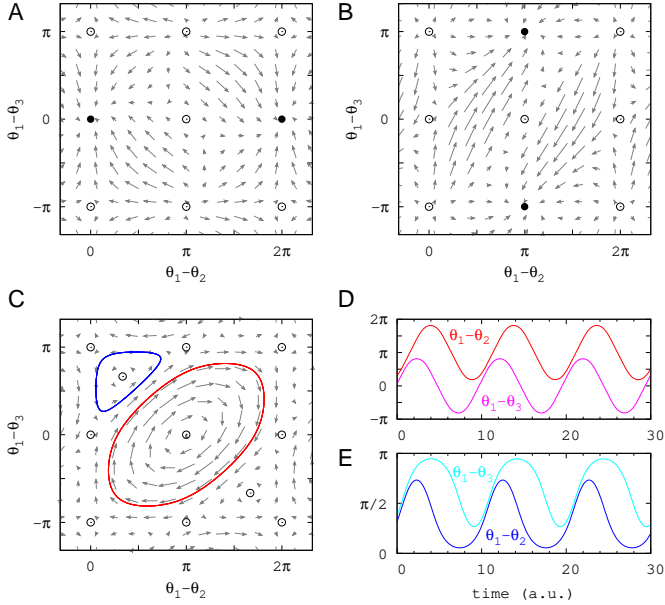}
\caption{(Color online) Vector flow (A) for the state $(0, 0)$ for 
$p  > q + r$;
(B) the state $(\pi, \pi)$ for $q  > p + r$;
and (C) traveling wave (limit-cycle) for $p = q = r$.
Note that filled/empty circles represent stable/unstable fixed points. 
The temporal evolution of $u(t)=\theta_1(t)-\theta_2(t)$ and 
$v(t)=\theta_1(t)-\theta_3(t)$ for the (D) red (clockwise limit cycle) and 
(E) blue (counter-clockwise limit cycle) regions in (C).
}
\label{fig:vectorflow}
\end{figure}
We find that the stability of the fixed point depends on  
the parameter values of $p$, $q$, and $r$: 
When the interaction by $\alpha$ cells is dominant 
($p > q+r$) at fasting conditions with low glucose levels, 
the system approaches the stable fixed point 
$(0, 0)$, showing 
an in-phase synchrony for $\alpha$-$\beta$, 
$\alpha$-$\delta$, and $\beta$-$\delta$ (Fig. ~\ref{fig:vectorflow}A). 

On the other hand, when the interaction by $\beta$ cells is dominant 
($q > p+r$) at fed conditions with high glucose levels, 
the system approaches the stable fixed point 
$(\pi, \pi)$, showing 
an out-of-phase synchrony for both 
$\alpha$-$\beta$ and $\alpha$-$\delta$, while an in-phase synchrony for 
$\beta$-$\delta$ (Fig.~\ref{fig:vectorflow}B). 
Note that when we additionally consider population densities,
the inequality ($q > p+r$) becomes $q n_2 > p n_1 + r n_3$.
Because most cells in the pancreatic islets are $\beta$ cells
($n_2 > n_1 >  n_3$), we naturally expect that 
the population dominance of $\beta$ cells 
is more likely to lead the islet system to the $(\pi, \pi)$ state.

At near normal glucose conditions when the dominance of 
$\alpha$ and $\beta$ cell interactions is relaxed (e.g., $p = q = r$),
present is a new stationary solution
\rev{of  limit cycles with $\dot u \neq 0$ and $\dot v \neq 0$.}
The limit cycles in the ($u$, $v$) 
plane (see Fig.~\ref{fig:vectorflow}C),
oscillate between the ($0$, $0$) and ($\pi$, $\pi$) states (see Figs.~\ref{fig:vectorflow}D and E).
\rev{We summarized these dynamic behaviors depending on relative coupling strengths
in the phase diagram of Fig. 3.}

What happens if $\delta$ cells are absent?
Biologically, this is a very important question since it may give some 
clue about the very reason why $\delta$ cells
are found in pancreatic islets.
According to our model, when $\delta$ cells are absent, 
Eq.~(\ref{eq:rho1_theta1}) and (\ref{eq:rho2_theta2}) are reduced to
\begin{equation}
\label{eq:u_without_delta}
\dot{u} = (q - p) \sin u.
\end{equation}
We find that $u=\pi$ is the stable fixed point for $q > p$, 
on the other hand $u=0$ is the stable fixed point for $p > q$. 
\rev{Note that for $p=q$, {\it traveling wave} (TW) states exist with $u \neq 0$ or $\pi$, but $\dot u = 0$.
This TW state has been reported in Ref.~\cite{ref:Hong_Strogatz_PRL};
it is known to be induced
by the asymmetry in the coupling parameters.   
In the islet system, the coupling is also asymmetric one 
$(J_{s^{\prime}s} \neq J_{s s^{\prime}})$, accordingly a TW state is 
naturally expected to appear.}
This implies that the change from one state (out-of-phase synchrony between the 
$\alpha$ and $\beta$ cells) to another 
one (in-phase synchrony between the cells) 
occurs drastically depending on the range of interaction strength.  
\rev{In the absence of $\delta$ cells, the drastic state change can be easily seen in the phase diagram
($r=0$) of Fig. 3.
Note that this is very awkward situation where small perturbations of glucose 
(influencing relative strengths of $p$ and $q$) can result in completely different states of islets.
In contrast, in the presence of $\delta$ cells, islets allow
flexible changes between $u=0$ and $u=\pi$ states using limit cycles
as shown for $r \neq 0$ in Fig. 3.
}

\begin{figure}
\includegraphics[width=7.5cm]{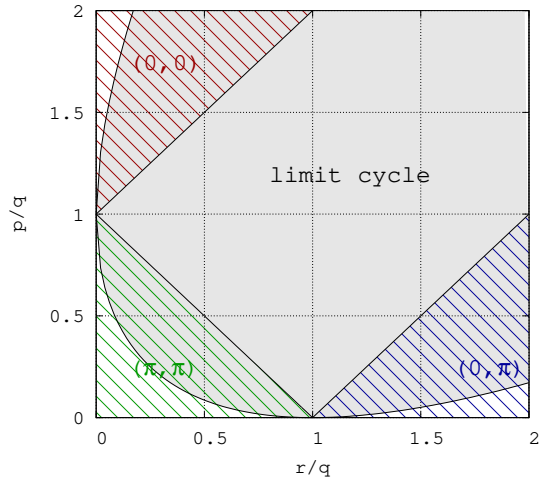}
\caption{\rev{
(Color online) Phase diagram for relative strengths of $p$, $q$, and $r$.
Stable fixed points ($\theta_1-\theta_2$, $\theta_1-\theta_3$)=(0, 0) for $p>q+r$ (red hatched area);
($\pi$, $\pi$) for $q>p+r$ (green hatched area); (0, $\pi$) for $r>p+q$ (blue hatched area); 
and limit cycles for $2pq+2pr+2qr-p^2-q^2-r^2>0$ (gray area).
Some regions have both fixed points and limit cycles as solutions depending on initial conditions.
}
}
\label{fig:phasediag}
\end{figure}

\section{Discussion}

In summary, we developed a model of coupled phase oscillators 
for the cells in pancreatic islets that explains 
their synchronized hormone pulses.
The model provides a clear picture about the 
characteristics of the cell-cell interactions in the islet, and 
suggests an important role of the third population, $\delta$ cells.

\rev{
In this paper, the islet system provides a natural extension of the Kuramoto model.
In the original model, every oscillator has the same positive coupling between them~\cite{ref:Kuramoto}.
Next simplest possible scenario may be to consider interactions between two distinct populations
in which one population (conformists) has positive coupling, while the other one (contrarians) has negative coupling.
This system has been demonstrated to show rich dynamics such as
out-of-phase synchrony between conformists and contrarians, and
traveling wave states where the phase difference between two populations is fixed, 
but each population still oscillates
with a new frequency different from their mean natural frequency~\cite{ref:Hong_Strogatz_PRL}.
Our study introduces a third population that is symmetrically distinct from the conformists and contrarians.
The third one should have a mixed coupling with positive and negative signs depending on neighbors.
The three-population model has larger flexibility in synchronization than the two-population model
as expected.
In addition to the in-phase and out-of-phase synchrony solutions, 
the limit cycle solutions allow three populations have periodic phase changes between them.
}

\rev{The islet system is an interesting realization of the three-population model.
Furthermore, the simple model of phase oscillators allows to understand 
biological meanings of symmetries of cellular interactions and functional roles of each population.
}
One main outcome of our model is the enlightenment of the
sign of the $\beta \rightarrow \delta$ interaction, $J_{23}$.
So far, consistency on the sign of interactions between $\alpha$, $\beta$, and $\delta$ cells
has been observed, except for the $J_{23}$ (see Table~\ref{table1}).
It has been reported that the interaction is positive in chicken pancreas~\cite{ref:J23a}, 
while other studies in canine (dog) pancreas reported it as
negligible~\cite{ref:J23b}. 
Although we can not exclude species differences,
the technical difficulty of measuring the infinitesimal amount of somatostatin ($\sim$femtomole),
may explain the inconsistency.
It has been reported that birds have surprisingly abundant $\delta$ cells in 
the islets, compared with mammal 
islets (40\% vs. 10\%)~\cite{ref:Hara} .
The extreme excess of $\delta$ cells in chicken might
allow to detect the stimulating effect of insulin secreted by $\beta$ cells.
In our model, we have found that if $J_{23} \leq 0$, it is impossible to generate
the reported in-phase synchronization between $\beta$ and $\delta$ cells.
{\color{black}{The}} positive interaction breaks the symmetry between 
$\beta$ and $\delta$ cells,
and gives three symmetrically-distinguishable cell populations (Fig. 1):
$\alpha$ cells activate other populations;
$\delta$ cells suppress other populations;
while $\beta$ cells stimulate and suppress other populations.
In other words, $\alpha$ cells are {\color{black}{only}} suppressed by 
other populations;
$\delta$ cells are {\color{black}{only}} activated by other populations;
while $\beta$ cells are {\color{black}{both}} activated and suppressed by other populations.
It is of interest that evolutionary lower species have only two reciprocal partners of $\alpha$ and $\beta$ cells,
while higher species are equipped with symmetrically different three cell populations~\cite{ref:Hara}.

In addition to the conjecture of $J_{23}>0$, we found a potential role of the third population, $\delta$ cells.
Regardless of the existence of $\delta$ cells, the 
islet model with an asymmetric interaction between $\alpha$ and $\beta$ cells 
produces both out-of-phase and in-phase hormone pulses
of $\alpha$ and $\beta$ cells depending on the dominance of the 
inhibitory (repulsive) interaction ($\beta \rightarrow \alpha$)
and the excitatory (attractive) interaction ($\alpha \rightarrow \beta$).
The different synchronization patterns may be beneficial for 
controlling glucose levels.
Under high glucose conditions, insulin plays a role to decrease glucose levels.
Continuous action of excess insulin can cause episodes of hypoglycemia (diminished glucose in blood),
which is more dangerous than hyperglycemia (excessive glucose in blood) 
because it results in shock and finally death.
Therefore, intermittent glucagon pulses at the high glucose conditions can 
prevent to enter into hypoglycemia.
If the glucagon pulses were in phase with insulin pulses,
their actions in the liver, increasing and decreasing 
blood glucose levels, would compete,
resulting in inefficient glucose control.
On the other hand, under low glucose conditions, insulin secretion becomes negligible,
remaining just at a basal insulin level, and glucagon plays a role to increase glucose levels.
The basal insulin helps cells in the body to absorb available glucose.
Therefore, at the low glucose conditions, the in-phase glucagon and insulin pulses 
can be beneficial,
because insulin accelerates the immediate absorption of glucose produced by glucagon. 
Indeed the out-of-phase state in a postprandial condition has been observed~\cite{ref:Menge},
and the in-phase state after an overnight fast has also been reported~\cite{ref:Lang}.
Then, one may wonder 
which states the islet takes at normal glucose levels. 
While the absence of $\delta$ cells allows 
only the two states of in-phase and out-of-phase,
the presence of $\delta$ cells generates 
an oscillating state between the two.
We suggest that this oscillation maximizes the flexibility of the islet system
to quickly respond to uncertain glucose inputs.
This last point is left for further study.

\rev{Finally, note that our simple phenomenological model is limited to explain
the physiological rationale for hormone pulsatility, 
although it has been proposed that the periodic exposure to the hormones can prevent
desensitization of their receptors, compared with their continuous exposure~\cite{ref:Hellman_review}.}

We thank Jean-Emile Bourgine for a critical reading of the manuscript.
This research was supported by Basic Science Research funded by NRF 
No. 2012R1A1A2003678 (H.H.), 
\rev{by Ministry of Science, ICT \& Future Planning No. 2013R1A1A1006655 (J.J),}
and by the Max Planck Society, the Korea Ministry of Education, Science and Technology,
Gyeongsangbuk-Do and Pohang City (J.J).


\begin{thebibliography}{20}
\bibitem{footnote1}
{The stability of the fixed point $(u^*, v^*)$ can be investigated by analyzing the Jacobian matrix {\color{black}{$J$ at that point}}, with the values of its trace and determinant: Note that $\rm{tr}(J)=\lambda_1+\lambda_2$, and $\rm{det}(J)=\lambda_1 \lambda_2$ for the two eigenvalues $\lambda_1$ and $\lambda_2$. Therefore, the stable solution can be determined from all negative eigenvalues with $\rm{tr}(J)<0$ and $\rm{det}(J)>0$.}
\end{thebibliography}
\end{document}